\documentclass[a4paper]{article}

\usepackage{spconf,amsmath,bm,graphicx,multirow,amssymb}
\usepackage{url}
\usepackage{hyperref}
\hypersetup{colorlinks=True, urlcolor=black}
\usepackage[numbers,sort&compress]{natbib}
\usepackage{booktabs}
\usepackage{multirow}
\usepackage{subfigure}
\usepackage{CJKutf8}

\ninept
\title{Improving Prosody Modelling with Cross-Utterance BERT Embeddings for End-to-end Speech Synthesis}
\name{Guanghui Xu, Wei Song, Zhengchen Zhang, Chao Zhang, Xiaodong He, Bowen Zhou
\thanks{Thanks to Xixin Wu, Guangzhi Sun, and Jing Huang for discussions.}} 
\address{JD AI Research}

\begin{document}

\maketitle
\begin{abstract}
Despite prosody is related to the linguistic information up to the discourse structure, most text-to-speech (TTS) systems only take into account that within each sentence, which makes it challenging when converting a paragraph of texts into natural and expressive speech. In this paper, we propose to use the text embeddings of the neighboring sentences to improve the prosody generation for each utterance of a paragraph in an end-to-end fashion without using any explicit prosody features. More specifically, cross-utterance (CU) context vectors, which are produced by an additional CU encoder based on the sentence embeddings extracted by a pre-trained BERT model, are used to augment the input of the Tacotron2 decoder. Two types of BERT embeddings are investigated, which leads to the use of different CU encoder structures. Experimental results on a Mandarin audiobook dataset and the LJ-Speech English audiobook dataset demonstrate the use of CU information can improve the naturalness and expressiveness of the synthesized speech. Subjective listening testing shows most of the participants prefer the voice generated using the CU encoder over that generated using standard Tacotron2. It is also found that the prosody can be controlled indirectly by changing the neighbouring sentences.

\end{abstract}


\section{Introduction}
Due to the advancement in deep learning, state-of-the-art text-to-speech (TTS) systems \cite{DBLP:conf/iclr/SoteloMKSKCB17,Wang:2017Tacotron,Shen:2018Tacotron2,DBLP:conf/iclr/PingPGAKNRM18,DBLP:conf/iclr/TaigmanWPN18} nowadays can generate high fidelity speech based on a single sentence without much difficulty. However, when synthesizing based on a paragraph of texts with multiple sentences, such as the scripts from an audiobook or a spoken dialogue, the resulted speech is often much less natural and expressive. 
This can be attributed to the lack of diversity in a set of factors, such as intonation, stress, and rhythm, 
which are collectively referred to as prosody and correlated with the style of speech. 

There is a long-standing interest in the modelling of prosody and speaking styles for TTS.
Regarding the statistical parametric speech synthesis (SPSS) approach \cite{Zen:2013}, styles are either pre-defined or automatically discovered, and can be modelled by individual transforms for hidden Markov models (HMMs)\cite{Eyben:2012,ChenLanZhou:2014}. Symbolic prosody labels \cite{Tesser:2013} are widely used as features for both HMM-based and deep-learning-based SPSS systems\cite{Zen:2013,Fan:2014,Liu:2016}. Regarding end-to-end speech synthesis based on the encoder-decoder models \cite{bengio:NMT,Sutskever:2014seq2seq},
the Tacotron structure \cite{Wang:2017Tacotron,Shen:2018Tacotron2} can be extended to include an additional encoder to learn prosody embeddings \cite{Skerry-Ryan:2018,wu2018rapid,LeeYounggun:2019}, which can be used to query a bank of randomly initialised global style tokens with an attention model to derive the style embeddings \cite{Wang:2018GST}. Alternatively, variational autoencoders \cite{Kingma:2014VAE} can be used to learn style or prosody embeddings 
to achieve a fine-grained control of the synthesized speech \cite{Henter:2018,Zhang:2019MS,Akuzawa1:2019,DBLP:conf/iclr/HsuZWZWWCJCSNP19,DBLP:conf/icassp/SunZWCZRRW20,DBLP:conf/icassp/SunZWCZW20}. 
More recently, efforts have been made to use additional text embeddings derived from pre-trained LMs, such as the BERT model \cite{DevlinCLT19}, to improve the modelling of prosody \cite{xiao2020improving,HayashiWTTTL19,abs-1906-07307}. 
Moreover, 
prosody modelling with texts consist of multiple sentences have also been studied for SPSS \cite{HuSZWHW16,AubinCWK19,farrus2016paragraph,watts2015sentence,zhu2018controlling,peiro2018paragraph}. Apart from sentence positions \cite{farrus2016paragraph,watts2015sentence,zhu2018controlling}, discourse relations (DRs), which describe the logical relationship between two discourse units like sentences, are also used to improve the prosody generation \cite{HuSZWHW16,AubinCWK19,peiro2018paragraph}.

In this paper, we propose to use the cross-utterance (CU) information to improve the generation of the prosody in the speech of the utterance in the end-to-end TTS framework.
A cross-utterance (CU) encoder is used to convert the text embeddings of the previous and future sentences into fixed-length CU context vectors.
To form the input vectors of the decoder model, the CU context vectors are concatenated with the corresponding output vectors from the standard phoneme encoder with respect to the current sentence.
Two types of sentence embeddings are derived from a pre-trained BERT language model (LM) using the chunked and paired sentence settings respectively \cite{DevlinCLT19}. 
A different CU encoder structure is proposed for each type of BERT embeddings, including the one with a modified
multi-head attention layer \cite{DBLP:conf/nips/VaswaniSPUJGKP17} used for the paired sentence setting. More specifically, in order to derive a different CU context vector for each phoneme, each output vector from the phoneme encoder is used to query the multi-head attention layer to obtain a different set of annotation vectors to integrate the BERT embeddings. 
Experimental results based on a Mandarin audiobook dataset and the LJ-Speech English audiobook dataset showed that the use of CU sentence embeddings can synthesize more natural and expressive speech and achieve more stable duration modelling. 
It was also found that an indirect control of the prosody can be achieved by varying the neighbouring sentences of the current one. 
Comparing to the previous studies \cite{HuSZWHW16,AubinCWK19,farrus2016paragraph,watts2015sentence,zhu2018controlling,peiro2018paragraph}, our approach is more ``end-to-end'' since it only uses the raw texts as input and does not require any explicitly defined symbolic prosody labels or DRs. 

The rest paper is organised as follows. Section~\ref{sec:review} reviews BERT and Tacotron2, and introduces the two types of CU embeddings derived from BERT. Section~\ref{sec:tts_CU} presents the structures of the CU encoders. Experimental setup and results are given in Section~\ref{sec:setup} and \ref{sec:results}. We conclude in Section \ref{sec:conclusions}.

\section{A Review of BERT and Tacotron2}
\label{sec:review}
\subsection{The BERT-derived sentence embeddings}
\label{BERT}
BERT is a method that learns general-purposed text embeddings based on a bi-directional Transformer encoder~\cite{DBLP:conf/nips/VaswaniSPUJGKP17}, which is trained to predict both randomly masked words and the next sentence in the multitask learning framework. 
A large amount of open-domain texts are used as the training data to obtain the pre-trained model. 
A special token \texttt{[CLS]} is added at the begin, and the final hidden state
corresponding to this token is used as the sentence embedding in the next sentence prediction task. In BERT training, since sentence sub-words are packed into chunks with a maximum token length of 512, another special token, \texttt{[SEP]}, is inserted at the boundary of each sentence to keep track of the original sentence structure. 

Two types of context embeddings can be derived from BERT: chunked sentence embedding (CSE) and paired sentence embedding (PSE). Let $u_{0}$ be the current sentence, $L$ be the number of sentences taken into account in the past or future context, for CSEs, 
we have
\begin{equation}\nonumber
    \begin{split}
        \bar{u}_{P}&=\{\texttt{[CLS]},u_{-L},\texttt{[SEP]},u_{-L+1},\texttt{[SEP]},\ldots,\texttt{[SEP]},u_{0}\} \\ \bar{u}_{N}&=\{\texttt{[CLS]},u_{0},\texttt{[SEP]},u_{1},\texttt{[SEP]},\ldots,\texttt{[SEP]},u_{L}\}
    \end{split}
\end{equation}
as the past and future chunks of sentences. The CSEs of $u_0$ are obtained at the \texttt{[CLS]} tokens of the $\bar{u}_{P}$ and $\bar{u}_{N}$. The BERT segment embeddings are set as one sentence. 
Alternatively, for PSEs, sentence pairs, such as $\{$\text{\texttt{[CLS]}}$,u_{L-1},\text{\texttt{[SEP]}},u_{L}\}$, can be used to derive the sentence embeddings 
at the \texttt{[CLS]} token of the each pair, 
since BERT was trained to predict the next sentence. The BERT segment embeddings for each word of the two sentences are set to the values of \texttt{A} and \texttt{B} correspondingly.
More details of BERT and the BERT-derived sentence embeddings can be found in \cite{DevlinCLT19}.

\subsection{Tacotron2}
Tacotron2 \cite{Shen:2018Tacotron2} is an end-to-end TTS approach based on the encoder-decoder model structure \cite{bengio:NMT}. It consists of two components: an encoder-decoder network that predicts the Mel-spectrograms from the input texts and a modified WaveNet which acts as a vocoder. The encoder of the spectrogram prediction network transfers the input text into character embeddings first, and then feeds the embeddings into $3$ convolutional layers. The output of the last convolutional layer was passed into a bi-directional long short-term memory (LSTM) layer, and the outputs of the LSTM layer are taken as the encoded features. In this work, we modify the encoder structure slightly, and augment the encoder features with CU BERT embeddings. The details are shown in Section~\ref{sec:tts_CU}. The decoder is a two-layer LSTM with the location sensitive attention mechanism \cite{bengio:NMT}. The output of the decoder is used to predict the mel spectrogram wiht a 5-layer covolutional post-net. Before passed through the LSTM layer, the prediction of last time step is fed into a pre-net. The stop tokens are also predicted during inference. 

\section{Cross-Uttrance Encoders}
\label{sec:tts_CU}
In this section, Two different CU encoder structures are explored to meet the characteristics of the two types of the BERT-derived sentence embeddings, CSE and PSE introduced in Section~\ref{BERT}. Regarding CSEs, the past and future embeddings are fused simply using a concatenation operation. Regarding PSEs, a multi-head attention layer is used to fuse the embeddings of all neighbouring sentences separately with respect to each input unit in the current sentence.

\subsection{Phoneme encoder for the current sentence}
In our implementation of the Tacotron2 structure, rather than directly using graphemes (which refer to the Chinese characters for Mandarin) as the input, an extra grapheme-to-phoneme (G2P) step is performed first to convert the graphemes into phonemes. 
Regarding Mandarin, tone labels are added to the phonemes, and the same phoneme with different tones are treated as different phonetic units. A 256-dimensional (-d) latent representation space is learned jointly with the rest of the TTS system to encode the phonemes. Therefore, we refer to the standard Tacotron2 encoder that encodes the phoneme sequence of the current sentence as the phoneme encoder. 

Let $p_1,p_2,\ldots,p_T$ be the phoneme sequence of the current sentence, $\bm{f}(p_t)$ is the $t$\,th 512-d output vector of the phoneme encoder. The phoneme embedding matrix $\mathbf{F}$ is defined as
\begin{equation*}
    \label{fmat}
    \mathbf{F}=[\bm{f}(p_1),\bm{f}(p_2),\ldots,\bm{f}(p_T)].
\end{equation*}

\subsection{CSE-based CU encoder}
\label{CSE}
To use CU information in the encoder-decoder structure, neighbouring sentence embeddings can be converted to form the CU context vectors, which augment the input to the decoder part of Tacotron2. 

\begin{figure}[t]
    \centering
    \includegraphics[scale=0.4]{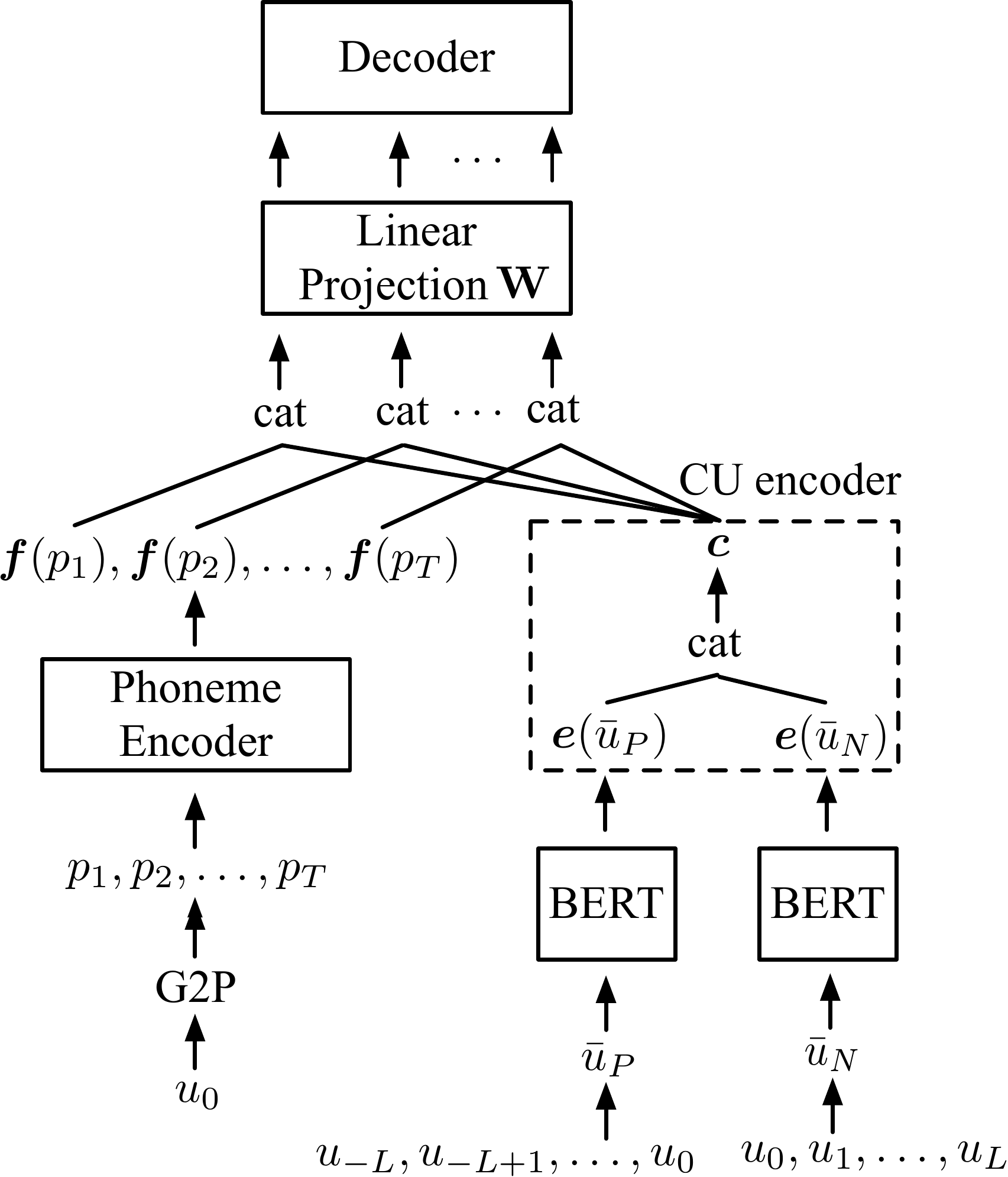}
    \caption{Model structure with the CSE-based CU encoder. The embeddings for the future and past chunked sentences are concatenated to form the CU context vector, which is concatenated with the phoneme encoder output vectors to form the input of the decoder.}
    \label{fig:packed_sentnece_structure}
    \vspace{-0.5cm}
\end{figure}

When using the CSEs to encode the CU information, 
two BERT embeddings, $\bm{e}(\bar{u}_P)$ and $\bm{e}(\bar{u}_N)$,
will be derived for each utterance $u_0$ based on its past and future sentence chunks, $\bar{u}_P$ and $\bar{u}_N$. The CU context vector $\bm{c}$ is defined as a simple concatenation of the BERT embeddings, $\bm{c}^{\text T}=[\bm{e}(\bar{u}_P)^{\text T},\bm{e}(\bar{u}_N)^{\text T}]$.
$\bm{c}$ is then concatenated with $\bm{f}(p_t)$ and transformed with a linear projection matrix $\mathbf{W}$ to form 
the 512-d input vector to the decoder at $t$, which helps to retain the same amount of parameters in the decoder model. Fig.~\ref{fig:packed_sentnece_structure} illustrates the network structure of the CSE-based CU encoder.

\subsection{PSE-based CU encoder}
\label{PSE}
PSEs can be used to construct CU encoders. 
For the current sentence $u_0$, the neighbouring sentences $u_{-L},u_{-L+1},\ldots,u_{L-1}$ can be used to derive $2L$ PSEs denoted as
\begin{equation*}
    \mathbf{E}=[\bm{e}(u_{-L},u_{-L+1}),\bm{e}(u_{-L+1},u_{-L+2}),\ldots,\bm{e}(u_{L-1},u_{L})].
\end{equation*}

To generate a CU context vector $\bm{c}_t$ for each input phoneme $p_t$, the multi-head attention layer from Transformer based on scaled dot-product \cite{DBLP:conf/nips/VaswaniSPUJGKP17} is used in the PSE-based CU encoder, where the value and the key are both set to the PSE matrix $\mathbf{E}$ and the query is set to the phoneme embedding matrix $\mathbf{F}$ produced by the phoneme encoder for $u_0$. Specifically, we have 
\begin{align*}
    [\bm{c}_1,\bm{c}_2,\ldots,\bm{c}_T]&=\text{Concat}(\bm{g}_1,\bm{g}_2,\ldots,\bm{g}_H)\mathbf{W}^{O},
\end{align*}
where $H$ is the number of heads in the attention model and $\text{Concat}(\cdot)$ is the concatenation operation. $\bm{g}_h$ is the annotation vector produced by the $h$\,th head that can be computed by
\begin{align*}
\bm{g}_h=\text{Attention}(\mathbf{F}\mathbf{W}^{Q}_h,\mathbf{E}\mathbf{W}^{K}_h,\mathbf{E}\mathbf{W}^{V}_h),
\end{align*}
where the attention function is defined by
\begin{align*}
\text{Attention}(\mathbf{Q},\mathbf{K},\mathbf{V})=\text{Softmax}(\mathbf{Q}\mathbf{K}^{\text T}/\sqrt{d_K})\mathbf{V}.
\end{align*}
Here $\mathbf{W}^{O}$, $\mathbf{W}^{Q}_h$, $\mathbf{W}^{K}_h$, and $\mathbf{W}^{V}_h$ are the linear projection matrices for output, query, key, and value, $d_K$ is the dimension of $K$, and $\text{Softmax}(\cdot)$ is the softmax activation function. Embeddings of the sentence indexes are added to the BERT-derived embeddings to mark their positions in the attention model.

\begin{figure}[t]
    \centering
    \includegraphics[scale=0.4]{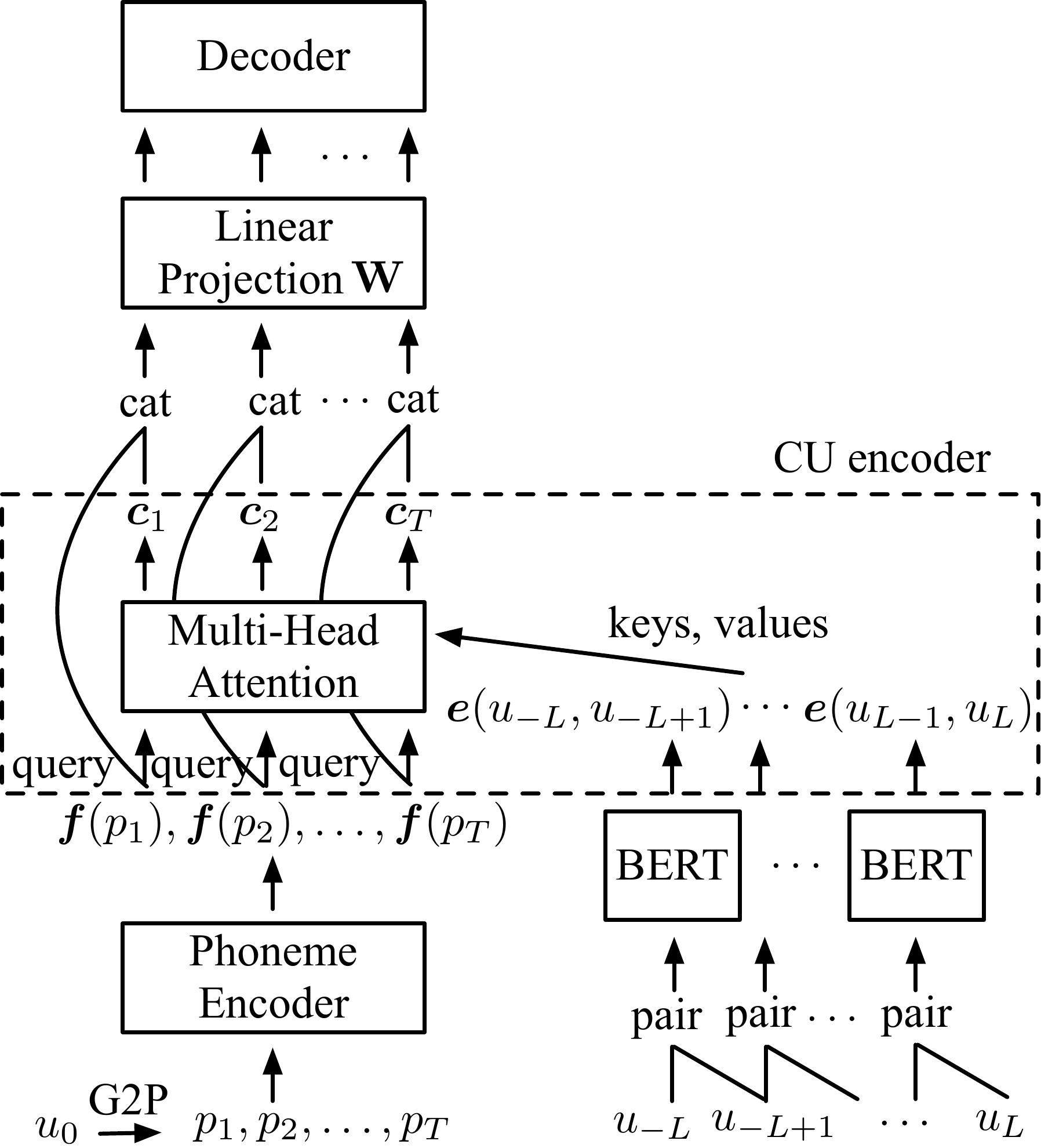}
    \caption{Model structure with the PSE-based CU encoder. A 
     paired sentence embedding is extracted for every sentence,  which are used to obtain the CU context vectors using 
     the multi-head attention layer.}
    \label{fig:cross_utterance_encoder}
    \vspace{-0.5cm}
\end{figure}

Similar to the CSE-based CU encoder, at every time step $t$, $\bm{c}_t$ is first concatenated with $\bm{f}(p_t)$ and then transformed by the linear projection $\mathbf{W}$ to form the 512-d input vector of the decoder.

\section{Experimental Setup}
\label{sec:setup}
An in-house Chinese Mandarin dataset and an English open dataset LJ-Speech~\cite{ljspeech17} were used in the experiments in this paper. Our Mandarin dataset has 29 hours of audio and 35k text-audio pairs. It was created by a professional male Mandarin native speaker reading a fiction novel with rich emotions. 
The speaking style varies from utterance to utterance, and the reading speed and the energy fluctuate even within a single sentence. Hence, it is very challenging to model this voice, and the standard Tacotron2 even could not achieve stable duration modelling. LJ-Speech contains 13.1k short audio clips created by a female native English speaker reading the scripts selected from 7 non-fiction books. The speaking style and the emotions were kept stable across the dataset, which makes it not as challenging and become a commonly used dataset for speech synthesis. 
We kept 100 pairs for validation, 100 pairs for testing and the rest for model training. All audio files were down-sampled to a 16kHz sample-rate, 
and the silence at both ends of every utterance is stripped. 

The pre-trained English and Chinese BERT-base models \cite{Wolf2019HuggingFacesTS} were used for our English and Mandarin experiments correspondingly, with each of them having 12 Transformer blocks and 12-head attention layers. 
The overall number of parameters in the BERT model is 110 million and the size of the derived embeddings are 768-d. 
No fine-tuning of the BERT model parameters was performed. 
An in-house G2P tool developed for our products was used to generate the phoneme sequences of the input text, and the sub-words segmentation tool was obtained from the BERT library. The 80-d Mel-spectrograms were used as the targets for training the model. The frame duration and shift are $50$ms and $12.5$ms respectively. 
The decoder predicts 2 frames at every step. 
 WaveRNN~\cite{KalchbrennerESN18} is used as the vocoder in each experiment. Unless mentioned specifically, all hyper-parameters were kept the same as used in  Tacotron2~\cite{Shen:2018Tacotron2}.

A subjective evaluation was conducted to verify the proposed method. Fifty native Mandarin speakers joined the listening tests in Mandarin, and 15 native English speakers joined those in English. Two evaluation metrics were used: the MUSHRA mean opinion score and the ABN choice preference rate~\cite{webmushra}. 
For ABN preference test, both in-domain (a fiction novel) and out-domain (introduction, prose and chat) a paragraph of scripts were used in the Mandarin evaluation. For our English models, the scripts were selected from the same non-fiction books for testing, which were not included in the training set.
Each utterance synthesized by the CU encoders used 2 sentences before and 2 sentences after the current script. All of the audio samples used in our tests can be found on our demo page, and we strongly encourage the readers to listen to them.\footnote{\url{https://xaviergithub.github.io/demos/CU_bert_embedding/index.html}}


\section{Experimental Results}
\label{sec:results}
\subsection{Influence of the BERT-derived sentence embeddings}
Four systems are compared in our experiments: 1) \textbf{Copy synthesis:} Copy synthesis from the ground truth Mel-spectrograms. 2) \textbf{Tacotron2:} The model built with the standard Tacotron2 approach. 3) \textbf{CSE-based model:} Tacotron2 with the CSE-based CU encoder. 4) \textbf{PSE-based model:} Tacotron2 with the PSE-based CU encoder. 
The subjective evalution results are shown in Table~\ref{tab:our_mos}. The PSE-based model outperforms the standard Tacotron2 on both datasets. The duration, pitch and stress of the speech produced by the PSE-model are obviously better than those produced by Tacotron2 when listening to the audios. The CSE-based model is slightly worse than Tacotron2 in terms of stability. Although the CSE-based model can generate the duration and pitch properly, there are some wrong tones in the test sounds. Meanwhile, we can see the scores of the English systems are generally lower than those of the Mandarin systems, which we believe is because the LJ-Speech audios are less emotional and expressive than those of our Mandarin dataset. The score of the standard Tacotron2 is close to that of the PSE-based model. This is due to the fact that the LJ-Speech audios are easy to learn even for the standard Tacotron2 which makes the advantages of our proposed methods less obvious.
\begin{table}[th]
    \centering
    \begin{tabular}{lcc}
    \toprule
    System & Mandarin                  & English                    \\
    \midrule
    Ground truth            & 89.2$\pm$2.4              & 94.1$\pm$1.7               \\
    Copy synthesis          & 90.6$\pm$2.1              & 87.8$\pm$2.2               \\
    \midrule
    Tacotron2              & 75.4$\pm$3.2              & 69.8$\pm$2.1               \\
    CSE-based model    & 72.8$\pm$3.3              & 67.3$\pm$3.0               \\
    PSE-based model   & \textbf{80.5$\pm$2.6}     & \textbf{70.7$\pm$2.4}      \\
    \bottomrule
    \end{tabular}
    \caption{The MUSHRA scores of the subjective evaluation. Fifty sentences selected from the testing set are evaluated. The means with 95\% confidence of the scores given by 15 subjects are listed.}
    \label{tab:our_mos}
    \vspace{-3mm}
\end{table}

\subsection{Comparisons between PSE- and CSE-based models}
Next, all listeners were requested to do the forced ABN test to compare the prosody 
generated by different systems, and the results are shown in Table~\ref{tab:open_domain}. In the Mandarin test, the subjects strongly prefer the PSE-based model comparing to either standard Tacotron2 or the CSE-based model, which reveals that the decoder can focus on the correct relevant CU information for each phoneme with the multi-head attention layer. The CSE-based model is the least preferred since it produced some incorrect tones. 
In the English test, both the PSE-based and CSE-based models are preferred by more listeners than the standard Tacotron2, while in Table~\ref{tab:our_mos} the MUSHRA scores of the three systems in English are more similar. 
 This is due to the fact that LJ-Speech is not challenging enough to demonstrate the superiority of our proposed methods, and the procedure of the ABN test makes the difference more obvious. 

\begin{table}[th]
    \centering
    \begin{tabular}{lccccc}
    \toprule 
    Language & ABN Choices  & Base          & CSE           & PSE           & Neutral       \\
    \midrule
     & Base vs. CSE      & \textbf{0.43} & 0.30          & -             & 0.27          \\
    Mandarin & Base vs. PSE      & 0.11          & -             & \textbf{0.73} & 0.16          \\
    & CSE vs. PSE       & -             & 0.10          & \textbf{0.69} & 0.21          \\
    \midrule
    & Base vs. CSE       & 0.34          & \textbf{0.50} & -             & 0.16          \\
    English & Base vs. PSE       & 0.38          & -             & \textbf{0.43} & 0.19          \\
    & CSE vs. PSE        & -             & \textbf{0.39} & 0.30          & 0.31          \\
    \bottomrule
    \end{tabular}
    \caption{The subjective listening preference rate, 
    where ``Base'', ``PSE'', and ``CSE'' refer to the standard Tacotron2, the PSE-based model, and the CSE-based model respectively.}
    \label{tab:open_domain}
    \vspace{-6mm}
\end{table}

\subsection{A Case study}
\begin{figure}[t]
    \centering
    \subfigure[Who called Mary? \textbf{Tom called Mary.}]{
    \label{fig:case_study:a} 
    \includegraphics[width=0.95\linewidth]{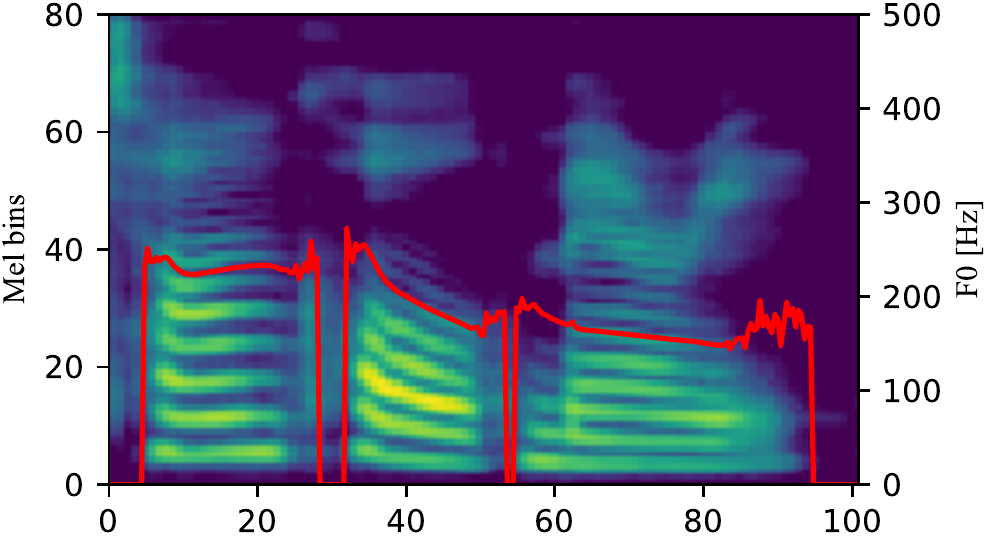} 
    } 
    \subfigure[What did Tom do with Mary? \textbf{Tom called Mary.}]{
    \label{fig:case_study:b} 
    \includegraphics[width=0.95\linewidth]{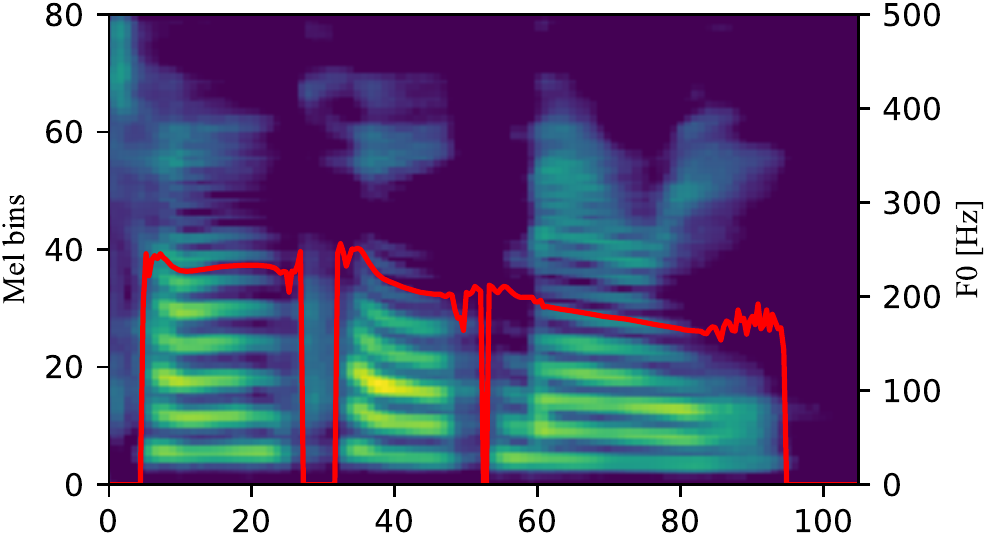}
    } 
    \caption{Comparisons between the Mel-spectrograms of ``Tom called Mary'' generated by Tacotron2 with the PSE-based CU encoder and different neighbouring sentence settings on LJ-Speech. Dropout is turned-off at test-time to avoid causing any random variations.}
    \label{fig:case_study}
    \vspace{-3mm}
\end{figure}
To investigate if the proposed methods can generate proper prosody to express the right meanings, a case study was made to synthesize the same words with different previous sentence as shown in Fig.~\ref{fig:case_study}. In the example, if the last sentence is the question ``Who called Mary?'', the key word in the reply ``Tom called Mary.'' is the subject ``Tom''. If the question is ``What did Tom do with Mary?'' then the key word in the same reply is the predicate ``called''. 
When our LJ-Speech PSE-based model was used to generate the examples, from Fig.~\ref{fig:case_study}, 
the two utterances have obviously different Mel-spectrograms. 
In particular, the fundamental frequencies have different changes. In Fig.~\ref{fig:case_study:a}, the speech launches with high pitch and the word ``Tom'' has longer duration, while in Fig.~\ref{fig:case_study:b}, the word ``called'' is emphasized and has higher pitch. 
The illustration of spectrograms match our expectation based on the linguistic knowledge and demonstrates that the PSE-based model can learn to use prosody to express the meanings in the right way. Note that dropout is turned-off at test-time in this case to avoid affecting our analysis. However, in the standard Tacotron \cite{Wang:2017Tacotron} and Tacotron2 \cite{Shen:2018Tacotron2} approaches, dropout is enabled at test-time to make random variations in prosody, which might result in incorrect expression of the meaning from the linguistic perspective. In addition, this example also reveals that an indirect control of the prosody generation can be achieved by varying the neighbouring sentences in our proposed methods. 


\section{Conclusions}
\label{sec:conclusions}
The use of the neighbouring sentence embeddings to improve the prosody modelling of TTS is proposed in this paper. Such CU information is encoded using an extra CU encoder added to Tacotron2.
Two types of the BERT embeddings are used, which leads to different model structures. 
Speech synthesis experimental results using a very challenging Mandarin audiobook dataset and using the commonly used LJ-Speech English dataset showed that the use of both types of BERT embeddings improved the MUSHRA scores or the ABN preference-rate over the standard Tacotron2. It is also demonstrated that the proposed methods can generate prosody to express the meaning in the expected way. 



\section{References}
{
\footnotesize
\begingroup
\renewcommand{\section}[2]{}
\bibliographystyle{IEEEtran}
\bibliography{mybib}
\endgroup
}




\end{document}